\documentstyle[nanosymp,epsf]{article}
\begin{document}

\title{Friedel oscillations of the magnetic field penetration in
systems with spatial quantization.}

\author{M. V. Entin,  L. I. Magarill, and M. M. Mahmoodian}


\affil{Institute of Semiconductor Physics, Siberian Branch of
the RAS, 630090 Novosibirsk, Russia}

\beginabstract

The magnetic field, applied to a size-quantized system produces
equilibrium persistent current non-uniformly distributed across
the system. The distributions of dia- and paramagnetic currents
and magnetic field in a quantum well is found. We discuss the
possibility of observation of field distribution by means of  NMR.

\endabstract

Traditionally, the magnetic field, penetrating into the system
with spatial quantization is considered as uniform and coinciding
with the external field. The magnetization of such system and
diamagnetic currents are weak and the corrections to the external
field are rather small. Nevertheless diamagnetic currents in
quantum systems are essential in such phenomena as NMR. It is well
known that the magnetic field acting on the atomic nucleus is
partially screened by electron shells that results in the chemical
shift of NMR line. This shift is measurable due to very narrow
width of NMR line as compared with the typical electron relaxation
rates. In a spatially quantized system non-uniform electronic
currents of magnetization also produce the screening of the
external field resulting in the change of effective magnetic field
acting on nuclei.

The orbital magnetism in systems with spatial quantization was
studied in a number of papers (see, e.g., \cite{Az}-\cite{Shap}).
These works consider the total magnetization of small systems. The
purpose of the present paper is to find the current and magnetic
field distribution in a quantum well.

\subsection*{Diamagnetic contribution.}

Let the magnetic field $\bf B$ to be directed along  $x$ axis
in the film plane $(x,y)$. The magnetic field in the film is
determined by the Maxwell equation $\partial {\bf B}/\partial
z=4\pi{\bf j}(z)/c $. The diamagnetic current density $\bf j$
has the only $y$ component. Since diamagnetism  is weak we
shall neglect
 corrections to the uniform external field in the expression
 for diamagnetic current. We shall consider diamagnetic current
 in linear in external magnetic field approximation.
 The equilibrium density of current can be found from the expression
\begin{equation}\label{jy}
  j_y(z)={\rm Sp}(\hat{j}_y(z)f(\hat{\cal H})),
\end{equation}
where $\hat{\cal H}=(\hat{\bf p}+e{\bf A}/c)^2/2m + U(z)  $ is
the electron Hamiltonian,   ${\bf A} = (0,-B_0z,0)$ is the
vector potential of the external magnetic field ${\bf B}_0$,
$U(z)$ is the confining potential, $ \hat{j}_y(z) =
-e\{\hat{v}_y,\delta(z-\hat{z})\}/S$ is the orbital current
density operator, $\hat{\bf v}=(\hat{\bf p}+e{\bf A }/c)/m$ is
the electron velocity operator, $\{...\}$ stand for the
operation of symmetrization,  \
$f(E)=(\exp{((E-\mu)/T)}+1)^{-1}$ is the Fermi function
($\mu,\ T$ are the chemical potential and the temperature),
$S$ is  the surface area of the system. Here and below
$\hbar=1$.

The linear approximation in $B_0$ yields:
\begin{equation}\label{jylin}
 j_y(z) = \frac{2e^2B_0}{mcS}\sum_{n,{\bf p}}\Bigl[(z-z_{nn})\varphi_n(z)^2
 f(E_{n,{\bf p}}) + \sum_{n'\neq n}\frac{p^2}{2m}\varphi_n(z)\varphi_{n'}(z)
z_{nn'}\frac{f_{n,{\bf p}}-f_{n',{\bf p}}}{E_{n,{\bf
p}}-E_{n',{\bf p}}}\Bigr]
\end{equation}
Here $\varphi_n(z)$ are the transversal wave functions, ${\bf
p}$ is the longitudinal momentum, $E_{n,{\bf p}}=E_{n}+p^2/2m$
is the energy of an electron in the  n-th subband of
transversal quantization, \ $f_{n,{\bf p}} \equiv f(E_{n,{\bf
p}})$.  In the model of a rectangular quantum well
("hard-wall" potential: $U(z)=0$ at $0<z<d, ~ U(z)=\infty$ for
$z<0$ and $z>d$) Eq. (\ref{jylin}) can be simplified for  $
T=0$ as
\begin{eqnarray}\label{jhw}
 j_y(E_F,0;z)=\frac{e^2
 B_0}{4
 \pi cd^2}\sum_{n=1}\frac{E_F-E_n}{E_n}\theta(E_F-E_n)\times\nonumber\\
 \Big[(E_F+3E_n)~d(2z-d)~ \sin^2{(\pi nz/d)}
 -(E_F-E_n)~n \pi z(z-d)~\sin{(2\pi nz/d)}\Big].
\end{eqnarray}
Here  $E_F=\mu(T=0)$ is the Fermi energy,
$E_n=\pi^2n^2/2md^2$.

The finite-temperature formula for current  can be obtained
from (\ref{jhw}) using the  relation
\begin{equation}\label{1}
  j_y(\mu,T;z)=\int dE~~j_y(E,0;z)(-\frac{\partial f(E)}{\partial
  E}).
\end{equation}

\begin{figure}[h]\centerline{\epsfbox{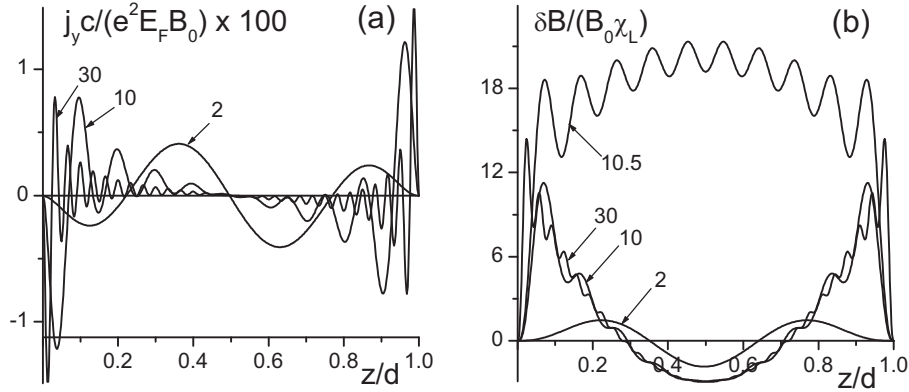}}
\leavevmode  \caption{ Evolution of the current density (a) and
magnetic field (b) for $T=0$ with the well width.  The numbers of
the occupied transversal subbands $k_Fd/\pi$ are marked on curves.
For $k_Fd/\pi=2,10,30$ the Fermi energy coincides  with subbands
bottoms, for $k_Fd/\pi=10.5$ it lies between 10th and 11th
bottoms.}
\end{figure}

The Fig. 1(a) depicts the current distribution in the quantum well
with rectangular walls. The current density is antisymmetric in
reference to the well center and oscillatory decays apart from the
boundaries. The oscillations reflects the Friedel phenomenon,
namely, the susceptibility singularity at the wave vector $2k_F$.
The direction of current density alternates, so the term
"diamagnetic", strictly speaking, refers to the integrated surface
current only. In the low-temperature limit the current decay is
slow. If the well width is larger than  the Fermi wavelength,
$2\pi/k_F$, the current density and magnetic field asymptotics on
the small distance $z\ll d$ to the boundary read
\begin{equation}\label{2}
j_y= 3\chi_Lck_FB_0\Bigl(-\frac{\cos
x}{x}+3x^2\int\limits_x^\infty\frac{\sin t}{t^5}dt\Bigr),~~~
\frac{\delta B}{ B_0} =4\pi\chi_{L}\Bigl( 1-\frac{3\sin
x}{2x}+\frac{3x^3}{2}\int\limits_x^\infty\frac{\sin
t}{t^5}dt\Bigr)
\end{equation}
Here $\chi_L=-e^2k_F/12\pi^2mc^2$ is the Landau diamagnetic
susceptibility at $T=0$, \ $x=2k_Fz$. The first terms in
Eq.(\ref{2})  represent the asymptotics in the domain $z\gg
\pi/k_F $. In particular, the constant contribution in $\delta
B/B_0$ gives exactly the Landau magnetic permeability of a bulk
sample.

At finite temperature one can find for $k_Fz\gg 1$
\begin{equation}\label{104}
\frac{\delta B}{B_0}=4\pi\chi_{L}\left[1-\frac{3\pi
T}{4E_F}\frac{\sin(2k_Fz)}{\sinh(z/l_T)}\right],
\end{equation}
where the  characteristic damping length $ l_T=k_F/(2\pi mT)$.
Note, that the  impurity scattering also leads to the drop of
$\delta B(z)$ on a distance of mean free path  from the surface.

Besides the oscillating surface diamagnetic current,  Fig. 1(a)
unexpectedly shows a small regular component of the current
density, which is linear with the transversal coordinate.
Asymptotically at $k_F d/\pi\gg1$ this contribution reads as
\begin{eqnarray}\label{4}
-\frac{e^2B_0k_F}{12mcd^2}(z-d/2)(1+6\zeta(\zeta-1)),~~~~~~~
\zeta=k_Fd/\pi-[k_Fd/\pi].
\end{eqnarray}
The linear component is smaller than the surface current by
the factor   $1/(k_F d)$. The sign of the slope of the linear
term alternates with the chemical potential. The linear term,
averaged with respect to the chemical potential (ensemble
averaging) vanishes. At the finite temperature the linear term
is
\begin{equation}\label{5}
-\frac{e^2TB_0}{\pi cd}(z-d/2)\sum\limits_{n=1}^\infty
\frac{\cos (2nk_Fd)}{n\sinh (nT/\theta)},
\end{equation}
where $\theta=E_F(\pi k_Fd)^{-1}$ is the characteristic
temperature above which the linear term washes out.

The magnetic field corrections are shown on the Fig. 1(b). The
linear term in $j_y$ produces the parabolic contribution to
magnetic field, which is sensitive to the parameter $\zeta$.  The
linear term in the current density and the parabolic contribution
to the magnetic field are connected with  the orbital magnetism.
In a quantum film the orbital contribution to the magnetic
susceptibility fluctuatively grows with the width like $k_Fd$,
that corresponds to the growth of the parabolic contribution to
the magnetic field.

\subsection*{Paramagnetic current}

In addition to the diamagnetic current, there is also a
paramagnetic current, caused by electron spins. This
contribution can be found  from Eq.(\ref{jy}) with taking into
account  Pauli part of the Hamiltonian $-g\mu_BB_0\sigma_x/2$
and spin-related component of the current density operator
$\hat{\bf
j}^{(sp)}=cg\mu_B~\nabla\times(\mbox{\boldmath{$\sigma$}}
\delta(z-\hat{z}))/S$. Here $g$ is g-factor, $\mu_B$ is the
Bohr magneton, $\sigma_i$ are Pauli matrices. As a result we
find for the density of paramagnetic current
\begin{eqnarray}\label{cv}
  j_y^{(sp)} = \frac{g^2\mu_B^2cB_0}{2S}\sum_{n,{\bf p}}\Bigl(-
  \frac{\partial f(E_{n{\bf p}})}{\partial E}\Bigr)\frac{\partial
  \varphi_n^2(z)}{\partial z}
  \equiv \frac{g^2\mu_B^2cB_0}{4}\frac{\partial}{\partial \mu}
  \frac{\partial}{\partial z}n(z),
\end{eqnarray}
where $n(z)$ is the local electron concentration. This current and
corresponding magnetic field should be added to the diamagnetic
contributions. The ratio of diamagnetic and paramagnetic
contributions depends on g-factor and, in principle, may strongly
vary in different materials.

In the specific case of square quantum well and $T=0$ we find
\begin{equation}\label{aa}
 j_y^{(sp)} =-\frac{g^2\mu_B^2mk_F^2c B_0}{2\pi^2}
 (\frac{\cos x}{x}-\frac{\sin x}{x^2}),~~~~
 \delta B=\frac{g^2\mu_B^2mk_F B_0}{\pi}(1-
 \frac{\sin x}{x}).
\end{equation}

Let us consider the imaginary experiment of excitation of nuclear
spin  transition by the alternating gate voltage. Let 2D gated
system is subjected to a magnetic field with z and x components.
The normal component of field stays unscreened in an infinite 2D
system.  The x-component of magnetic field depends on the number
of electrons and their state and hence can be controlled by acting
on electron subsystem. In particular, the alternating voltage,
applied to the gate will modulate the magnetic field and excite
the NMR transitions. The excitation of NMR transitions is possible
if the alternating component of magnetic field is orthogonal to
the constant magnetic field.  The resonance should be detected by
the frequency (or magnetic field) dependence of the gate
impedance. The magnetic field is non-uniform and hence different
nuclei experience different in value magnetic fields. On the one
hand, this  leads to inhomogeneous signal broadening, on the other
hand,  provides the way of separate excitation of nuclei at
different depths.

The work was supported by the Russian Fund for Basic
Researches (grants  00-02-17658 and 02-02-16377).

\end{document}